\documentclass{article}

\usepackage{PRIMEarxiv}

\usepackage[utf8]{inputenc} 
\usepackage[T1]{fontenc}    
\usepackage{hyperref}       
\usepackage{url}            
\usepackage{booktabs}       
\usepackage{amsfonts}       
\usepackage{nicefrac}       
\usepackage{microtype}      
\usepackage{fancyhdr}       
\usepackage{graphicx}       
\usepackage{amsmath}        
\usepackage{multirow}       
\graphicspath{{media/}}     

\title{Let Distortion Guide Restoration (DGR): A Physics-Informed Learning Framework for Prostate Diffusion MRI}

\author{
  \textbf{Ziyang Long$^{1,2}$, Binesh Nader$^{3}$, Lixia Wang$^{1}$, Archana Vadiraj Malaji$^{1}$} \\
  \textbf{Chia-Chi Yang$^{1}$, Haoran Sun$^{1,2}$, Rola Saouaf$^{3}$, Timothy Daskivich$^{4}$} \\
  \textbf{Hyung Kim$^{4}$, Yibin Xie$^{1}$, Debiao Li$^{1,2}$, Hsin-Jung Yang$^{1}$} \\[8pt]
  $^{1}$Biomedical Imaging Research Institute, Cedars-Sinai Medical Center, Los Angeles, CA, USA \\
  $^{2}$Department of Bioengineering, University of California, Los Angeles, CA, USA \\
  $^{3}$Imaging Center, Cedars-Sinai Medical Center, Los Angeles, CA, USA \\
  $^{4}$Urology, Cedars-Sinai Medical Center, Los Angeles, CA, USA \\[4pt]
  $^{*}$Corresponding author: \texttt{Hsin-Jung.Yang@cshs.org}
}

\begin{document}
\maketitle

\begin{abstract}
We present Distortion-Guided Restoration (DGR), a physics-informed hybrid CNN--diffusion framework for acquisition-free correction of severe susceptibility-induced distortions in prostate ssEPI DWI. DGR is trained to invert a realistic forward distortion model using large-scale paired distorted/undistorted data synthesized from distortion-free prostate DWI and co-registered T2W images from 410 multi-institutional studies, together with 11 measured B$_0$ field maps from metal-implant cases incorporated into a forward simulator to generate low-b DWI (b = 50 s/mm$^2$), high-b DWI (b = 1400 s/mm$^2$), and ADC distortions. The network couples a CNN-based geometric correction module with conditional diffusion refinement under T2W anatomical guidance. On a held-out synthetic validation set (n = 34) using ground-truth simulated distortion fields, DGR achieved higher PSNR and lower NMSE than FSL TOPUP and FUGUE. In 34 real clinical studies with severe distortion, including hip prostheses and marked rectal distension, DGR improved geometric fidelity and increased radiologist-rated image quality and diagnostic confidence. Overall, learning the inverse of a physically simulated forward process provides a practical alternative to acquisition-dependent distortion-correction pipelines for prostate DWI.
\end{abstract}

\keywords{Prostate MRI \and Diffusion-weighted imaging \and Susceptibility distortion \and Deep learning \and Physics-informed neural network}

\section{Introduction}

Multiparametric MRI (mpMRI), combining T$_2$W imaging with multi-b-value DWI, is central to prostate cancer detection, PI-RADS grading, and treatment planning~\cite{antunes2021susceptibility}. High-b DWI and the derived ADC maps provide key biomarkers of lesion conspicuity and tumor aggressiveness, yet they are intrinsically low in signal-to-noise ratio (SNR) and susceptible to imaging artifacts. Even moderate distortions can blur lesion boundaries, shift anatomical landmarks, or corrupt ADC estimates, which directly impacting diagnostic and clinical decisions. As a result, susceptibility-induced distortion has become one of the most persistent technical barriers to high-quality prostate mpMRI.

These challenges are greatly amplified in patients with bowel distention and metallic implants (e.g., unilateral or bilateral hip prostheses)~\cite{hargreaves2011metal,koch2010magnetic}. Susceptibility perturbations from metal extend into the pelvis, generating B$_0$ gradients far exceeding the effective EPI bandwidth. As a result, ssEPI (the clinical standard for prostate DWI) images are susceptible to extreme stretching, compression, pixel pile-up, and complete signal dropout within the gland. Such distortions often eliminate high-b DWI and ADC contrast to the point of non-diagnostic quality, a limitation repeatedly noted in clinical radiology reports. With hip arthroplasty increasingly common in the same population at highest risk for prostate cancer, robust distortion correction has become an urgent and still unresolved clinical need~\cite{kurtz2007projections}.

Classical correction techniques require an additional acquisition of the B$_0$ field-map (e.g., FMRIB's Utility for Geometrically Unwarping EPI (FUGUE), part of the FMRIB Software Library (FSL)), and reverse phase-encoding (PE) reconstruction (e.g., Tool for Estimating and Correcting Susceptibility-Induced Distortions (FSL TOPUP)). These methods can perform adequately when distortion is mild and the underlying signal remains intact~\cite{andersson2003correct,jezzard1995correction,ruthotto2012diffeomorphic,kybic2000unwarping}. However, all rely on accurate estimation of pixel displacement or local image gradients. Near metal implants, these assumptions break down. Severe signal dropout invalidates the forward signal model, large B$_0$ gradients invalidate linear approximations, and registration becomes ill-posed when anatomical structures are stretched or collapsed beyond recovery. Moreover, additional acquisitions such as dual PE images or reliable B$_0$ maps are often impractical in routine prostate MRI and may themselves fail in regions of severe susceptibility.

More recently, deep learning approaches have been explored to improve diffusion-weighted prostate imaging and mitigate metal-related artifacts. Vendor-specific reconstruction networks and post-processing models can enhance SNR and perceived image quality, and physics-informed or attention-guided networks have shown promise for multi-shot DWI reconstruction or metal artifact reduction~\cite{chen2023deep,ueda2022deep,qian2022physics,kim2019attention}. Yet these methods typically target denoising or generic artifact suppression, require specialized acquisitions (e.g., multi-shot, dual-polarity readouts), or address different anatomical sites and artifact mechanisms~\cite{qiu2025unsupervised}. None directly learns to undo the extreme geometric warping of ssEPI prostate DWI caused by hip prostheses or large rectum distention using only routinely acquired clinical data.

To overcome these limitations, we follow a physics-informed learning paradigm: simulate artifacts through medical physics and learn the inverse through neural networks~\cite{qian2022physics}. Rather than relying on real paired artifact-free data, we reframe distortion correction as learning the inverse of a forward ssEPI distortion process. A forward physical simulator generates corrupted images, and the network learns to restore anatomy from these physics-driven artifacts. This ``Let Distortion Guide Restoration'' formulation preserves the interpretability of physics-based modeling while allowing a neural network to learn robust inverse mappings that remain stable under extreme pixel displacement or signal loss.

\section{Material and Method}

\subsection{Dataset}

This study used two complementary prostate multiparametric MRI cohorts: the fastMRI Prostate dataset, which contains 314 studies, and an in-house Cedars--Sinai cohort of 130 prostate MRI examinations (Table~\ref{tab:dataset}). To drive the forward ssEPI simulator, we additionally collected 11 B$_0$ field maps from prostate MRI studies of patients with unilateral or bilateral hip prostheses. Among these 444 studies, 34 cases (5 fastMRI and 29 Cedars--Sinai) showed visually severe geometric distortion on the DWI images. These 34 subjects were reserved exclusively as the clinical test set and were never used for forward-model simulation or reverse learning network training.

The remaining 410 distortion-free studies (309 fastMRI and 101 Cedars--Sinai) formed the synthetic training pool for generating paired distorted/undistorted data. For this pool, low-b DWI, high-b DWI, ADC maps, and co-registered T$_2$-weighted images were used as undistorted references. From these 410 subjects, 34 were randomly selected and held out as a synthetic validation subset for quantitative evaluation, while the other 376 subjects were used for training the restoration network. Thus, no T$_2$W/DWI volumes from the 34 clinical test subjects contributed to the simulated training pairs, avoiding subject-level leakage between training/validation and test.

\begin{table}[!htbp]
  \centering
  \begin{tabular}{ccccc}
    \toprule
    Name & Scanner & T2W & DWI(ss-EPI) & Subject \\
    \midrule
    \multirow{2}{*}{FastMRI Prostate Dataset} & \multirow{2}{*}{Siemens Vida} & TR(s): 3.5--7.2 & TR(s): 5.0--7.3 & \multirow{2}{*}{314$^{*}$} \\
                             &              & TE(ms): 100     & TE(ms): 77      &           \\
    \midrule
    \multirow{2}{*}{In-House Prostate Dataset} & \multirow{2}{*}{Siemens mMR} & TR(s): 4.0--5.5 & TR(s): 4.5--5.0 & \multirow{2}{*}{130$^{\wedge}$} \\
                              &             & TE(ms): 80--132 & TE(ms): 73--95  &                \\
    \midrule
    Sum &  &  &  & 444 \\
    \bottomrule
  \end{tabular}
  \caption{\textbf{Summary of prostate mpMRI datasets used in this study.} Acquisition parameters for T$_2$W and ssEPI DWI sequences are listed for the fastMRI Prostate dataset and the Cedars--Sinai cohort. A total of 444 subjects were included. $^{*}$Among them, five subjects with severe distortion comprised the test set. $^{\wedge}$Of these, 29 subjects with severe distortion were used as the test set.}
  \label{tab:dataset}
\end{table}

\subsection{Preprocessing and Forward Distortion Simulation}

All mpMRI sequences from each study were rigidly registered and resampled to a common 256$\times$256 in-plane matrix with 0.56$\times$0.56 mm$^2$ pixel spacing, while preserving the original through-plane resolution. The B$_0$ field maps described in the Dataset section were denoised and band-limited using a spherical-harmonics filter to reduce scanner-specific noise and remove non-physiologic high-frequency components. Low-order components preserving bulk field structure were retained, while higher-order components were perturbed in a controlled manner to broaden distortion diversity. For each 2D B$_0$ slice, a voxel displacement map (VDM) was computed using the standard ssEPI formulation~\cite{jezzard1995correction}:
\begin{equation}
  \text{VDM} = S_{pe} \cdot \Delta f(r) \cdot \frac{N_{pe} \cdot \frac{PF}{R} - 1}{ESP}
\end{equation}
where $S_{pe}$ is the PE direction (+1/--1), $\Delta f(r)$ is the off-resonance at pixel $r$, $N_{pe}$ is the number of PE lines, $PF$ is the partial-Fourier factor, $R$ is the in-plane acceleration factor, and $ESP$ is the echo spacing. These VDMs were applied to distortion-free DWIs by displacing and splatting voxel intensities along the phase-encoding direction according to the prescribed shifts, thereby reproducing stretching, compression, pixel pile-up, and local signal dropout characteristic of ssEPI distortion.

\begin{figure}[!htbp]
  \centering
  \includegraphics[width=\textwidth]{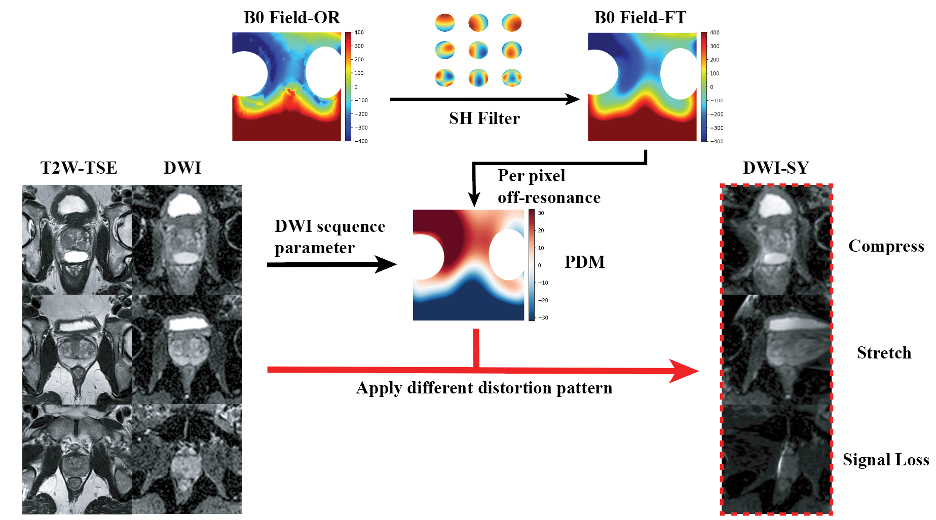}
  \caption{\textbf{Physics-based forward distortion simulation for ssEPI.} Non-distorted prostate DWIs (DWI) and T$_2$W (T2W-TSE) images serve as anatomical references. B$_0$ field maps from hip-prosthesis patients (B$_0$ Field-OR) are denoised and low-pass filtered using a spherical-harmonics (SH) model to generate smooth field variants (B0 Field-FT). Per-pixel off-resonance values are converted into voxel-displacement maps (PDMs) according to the ssEPI forward model and applied to DWI using a splat-based EPI simulator. The synthesized distorted DWIs (DWI-SY) naturally exhibit a spectrum of characteristic artifacts---including compression, stretching, pixel pile-up, and local signal loss---providing paired distorted/undistorted data for supervised training.}
  \label{fig:forward_sim}
\end{figure}

Using this pipeline (Figure~\ref{fig:forward_sim}), we generated more than 40,000 paired distorted/undistorted images from the 410 training-pool subjects, incorporating multiple phase-encoding directions (left--right, right--left, anterior--posterior, and posterior--anterior). The synthesized dataset captured a wide spectrum of realistic distortion patterns representative of metal-affected pelvic imaging and served as the primary supervision for training the inverse (restoration) networks.

\subsection{Network Architecture}

\begin{figure}[!htbp]
  \centering
  \includegraphics[width=\textwidth]{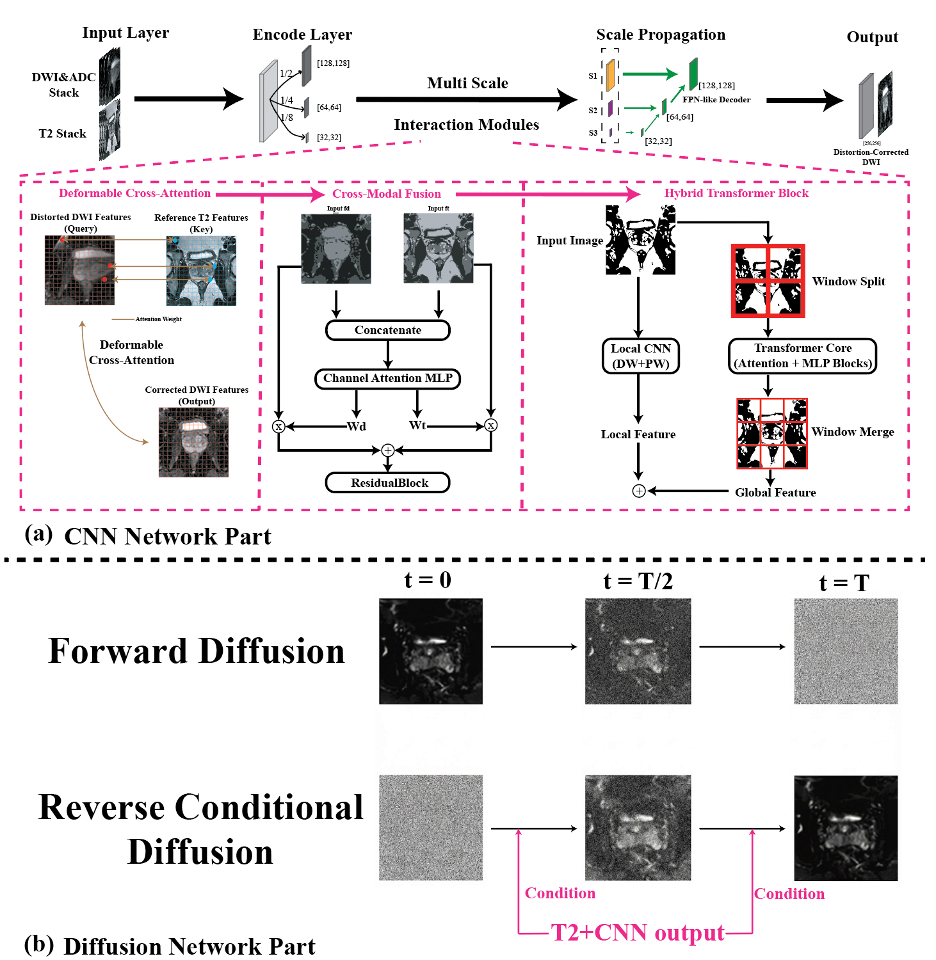}
  \caption{\textbf{Overview of the proposed inverse restoration network.} (a) CNN front-end: A multi-scale encoder with deformable cross-attention, cross-modal fusion, and hybrid transformer blocks extracts geometry-aware features from distorted DWI/ADC inputs under T2-weighted anatomical guidance, producing coarse distortion-corrected outputs. (b) Conditional diffusion refinement: During the forward diffusion process, clean images are progressively noised; the reverse conditional process denoises the CNN prediction toward high-fidelity results using T2+CNN features as conditioning signals. This two-stage design enables physics-guided coarse inversion and diffusion-based texture and detail restoration.}
  \label{fig:network}
\end{figure}

Building upon the forward EPI distortion simulator described above, we designed a physics-informed restoration network (DGR) using a hybrid CNN--diffusion architecture to approximate the inverse of the simulated distortion (Figure~\ref{fig:network}). Given distorted low-b DWI and ADC maps, along with an aligned T$_2$W image, the network predicts corrected low-b DWI and ADC maps and calculates the high-b DWI from them.

The CNN front-end operates as a coarse, physics-aligned inverse model. Stacks of distorted low-b DWIs and ADC are processed jointly with the registered T2W image by a compact multi-scale encoder--decoder~\cite{he2016deep}. The CNN aggregates geometry-aware features across scales and uses T2W-conditioned attention to encourage anatomical consistency between the corrected diffusion contrast and the reference anatomy~\cite{dai2017deformable,lin2017feature}. This branch produces initial low-b and ADC predictions that already remove most bulk distortion but remain overly smooth in severely corrupted regions, particularly on ADC where realistic noise patterns are hard to reproduce.

To restore fine texture and more natural signal statistics, we add a conditional diffusion refinement module that operates on the CNN outputs rather than generating images from pure noise~\cite{saharia2022image}. During training, a diffusion UNet receives a noisy version of the undistorted low-b/ADC pair together with the frozen CNN predictions and the co-registered T$_2$W image as conditioning. At inference, the process is initialized from the CNN output and perturbed with a small amount of noise (SDEdit-style), allowing the diffusion model to sharpen edges and reinstate realistic noise behavior while preserving the overall anatomy provided by the CNN and T$_2$W~\cite{meng2021sdedit}. The CNN is supervised with a weighted L1 loss, whereas the diffusion stage follows a standard DDPM-style clean-image prediction objective. Detailed architectural configurations and training schedules are provided in the Supplementary Methods.

\subsection{Evaluation}

The proposed method DGR was evaluated using both synthetic and clinically acquired datasets.

Synthetic datasets (34 subjects): Quantitative evaluation used PSNR and NMSE computed between corrected images and their undistorted ground-truth references. In addition, for evaluation only, a reverse phase-encoded distorted counterpart was generated to enable FSL TOPUP to run under its required dual-encoding input.

For comparison with classical correction pipelines, both FSL FUGUE and FSL TOPUP were configured under idealized conditions. For FUGUE, the ground-truth field map produced by the forward simulator was supplied directly, bypassing the noise, phase-unwrapping artifacts, and sequence mismatch that typically degrade field-map--based correction in vivo. Thus, the reported FUGUE performance represents an upper bound rather than typical clinical performance. Similarly, TOPUP was provided with the ground-truth distortion field in place of its own estimated field map, isolating its reconstruction behavior from field-estimation uncertainty and representing the theoretical limit of reverse phase-encoding correction. The default TOPUP configuration, in which the field map is estimated from the forward/reverse ssEPI pair, was also tested; its performance is reported in the Results section.

Clinical datasets (34 subjects): Qualitative evaluation was performed on the clinical test cohort, where undistorted DWI is not obtainable in vivo. Therefore, co-registered T2-weighted images were used as a soft anatomical reference. The model received the distorted low-b DWI, ADC map, and T$_2$W as inputs. Two abdominal radiologists independently scored each dataset on three separate 5-point Likert scales (1=poor, 5=excellent) for geometric fidelity, overall image quality, and diagnostic confidence.

\section{Results}

\subsection{Synthetic Experiments with Ground-Truth Undistorted DWIs}

Quantitative results are summarized in Table~\ref{tab:results}. Across both low-b (b = 50 s/mm$^2$) DWI and ADC maps, the proposed DGR model achieved the highest PSNR and the lowest NMSE among all methods. Specifically, DGR reached 23.88$\pm$2.93 dB / 0.089$\pm$0.049 (PSNR / NMSE) for low-b DWI and 22.99$\pm$1.97 dB / 0.062$\pm$0.028 for ADC, outperforming both FUGUE and TOPUP approaches. The full DGR pipeline required 13 to 15 seconds per subject on an NVIDIA HGX H100 GPU.

For reference, when TOPUP relied on its own estimated field map rather than the ideal situation, NMSE increased to 0.227$\pm$0.196 and 0.193$\pm$0.121 for low-b DWI and ADC respectively, underscoring the difficulty of robust field estimation in highly distorted pelvic imaging.

\begin{table}[!htbp]
  \centering
  \begin{tabular}{cccc}
    \toprule
    Method & Contrast & PSNR(dB)$\uparrow$ & NMSE$\downarrow$ \\
    \midrule
    \multirow{2}{*}{Baseline} & b=50 & 17.67$\pm$4.06 & 0.447$\pm$0.355 \\
             & ADC  & 15.60$\pm$2.51 & 0.351$\pm$0.172 \\
    \midrule
    \multirow{2}{*}{FUGUE + Fieldmap$^{*}$} & b=50 & 17.92$\pm$7.11 & 0.903$\pm$1.334 \\
                           & ADC  & 22.37$\pm$2.51 & 0.073$\pm$0.040 \\
    \midrule
    \multirow{2}{*}{TOPUP + Rev PE$^{*}$} & b=50 & 22.45$\pm$4.83 & 0.175$\pm$0.181 \\
                         & ADC  & 19.42$\pm$2.92 & 0.157$\pm$0.107 \\
    \midrule
    \multirow{2}{*}{DGR} & b=50 & \textbf{23.88$\pm$2.93} & \textbf{0.089$\pm$0.049} \\
        & ADC  & \textbf{22.99$\pm$1.97} & \textbf{0.062$\pm$0.028} \\
    \bottomrule
  \end{tabular}
  \caption{\textbf{Quantitative comparison of distortion-correction methods on the synthetic dataset.} PSNR and NMSE were computed between the restored images and the undistorted reference for both low-b (b = 50 s/mm$^2$) DWIs and ADC maps. DGR achieved the highest PSNR and lowest NMSE across both contrasts. $^{*}$FUGUE and TOPUP are evaluated under an ideal setting that requires additional acquisitions (dual-echo GRE field maps or reverse phase-encoding DWI pairs) and high-quality distortion fields. DGR operates on the routine mpMRI protocol without extra acquisitions.}
  \label{tab:results}
\end{table}

\begin{figure}[!htbp]
  \centering
  \includegraphics[width=\textwidth]{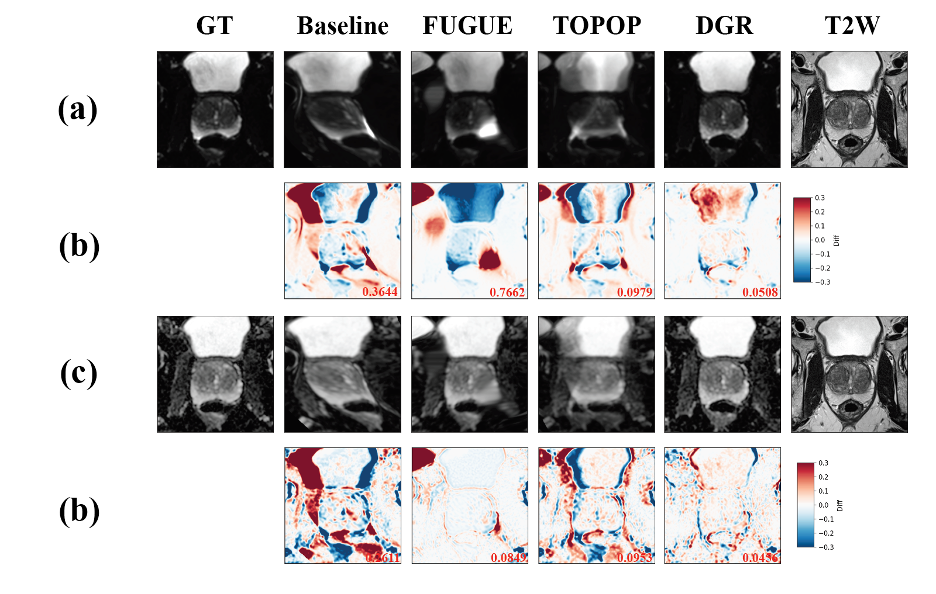}
  \caption{\textbf{Representative synthetic example comparing distortion-correction methods.} (a) Corrected low-b (b = 50 s/mm$^2$) DWIs produced by each method, with the undistorted ground truth (GT) and co-registered T2W shown for reference. (b) Corresponding voxel-wise difference maps (DWI -- GT), with normalized mean squared error (NMSE) labeled. (c) Corrected ADC maps and their difference maps (bottom row). The proposed DGR method achieves substantially lower reconstruction error and improved geometric fidelity compared with the baseline, FUGUE, and TOPUP.}
  \label{fig:synthetic_results}
\end{figure}

A representative synthetic example is shown in Figure~\ref{fig:synthetic_results}. DGR produces markedly improved geometric alignment relative to the baseline, FUGUE, and TOPUP. The error maps further quantify these gains: for b = 50 DWI, DGR achieves the lowest voxel-wise error (0.0508 vs.\ 0.3644, 0.7662, and 0.0979), and similarly for ADC (0.0436 vs.\ 0.2761, 0.0849, and 0.0935). Taking undistorted GT and co-registered T$_2$W as reference, conventional approaches exhibit distinct failure modes. FUGUE introduces an unnatural hyperintense patch in the right inferior peripheral zone on the low-b DWI, likely due to inaccurate pixel remapping and Jacobian modulation under severe geometric deformation. The same region becomes excessively smooth on the ADC map, erasing fine texture that is essential for quantitative interpretation. TOPUP avoids such extreme localized artifacts but produces globally blurred reconstructions: both low-b DWI and ADC appear veil-like, with softened boundaries and loss of granular structural detail, together with mild residual pile-up tendencies. In contrast, DGR restores the prostate geometry with capsule curvature and internal glandular structures that closely match the undistorted GT and T$_2$W reference. No abnormal brightening or blurring is observed, and the internal signal texture remains anatomically plausible rather than oversmoothed.

\subsection{Clinical Experiments}

For clinical prostate DWI scans, no undistorted imaging ground truth is available, so evaluation focused on how DGR behaved on severely degraded images.

\begin{figure}[!htbp]
  \centering
  \includegraphics[width=\textwidth]{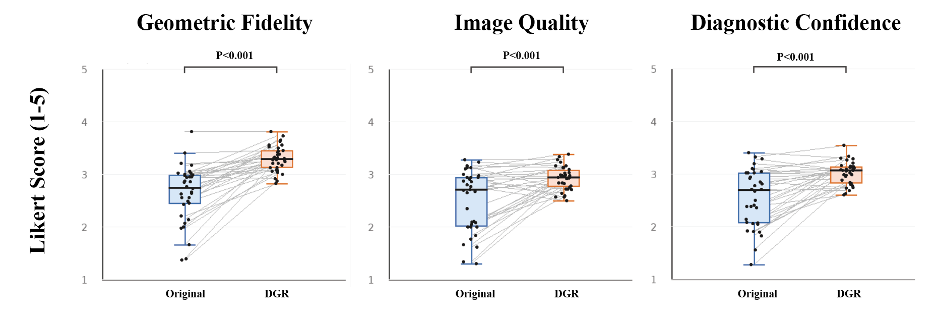}
  \caption{\textbf{Radiologist scores for original versus DGR-processed prostate diffusion images.} Boxplots show 5-point Likert ratings (1 = poor, 5 = excellent) for (left) Geometric Fidelity, (middle) Image Quality, and (right) Diagnostic Confidence in 34 clinical ssEPI studies. For each subject, readers scored the original diffusion images and the corresponding DGR-corrected images. Black dots represent per-subject averages across all evaluated slices, and gray lines connect paired scores from the same subject to visualize individual improvements. P-values from paired t-tests demonstrate significant improvements with DGR for all three metrics.}
  \label{fig:radiologist_scores}
\end{figure}

At the group level, DGR improved reader-perceived image quality and diagnostic usability across the 34 clinical subjects with susceptibility-induced distortion (Figure~\ref{fig:radiologist_scores}). Blinded radiologists assigned systematically higher scores to DGR-processed images than to the original ssEPI, with mean geometric fidelity increasing from 2.6 to 3.3, overall image quality from 2.5 to 2.9, and diagnostic confidence from 2.5 to 3.0 on a 5-point Likert scale (all $p < 0.001$, paired t-tests). The consistent upward shifts in paired subject scores indicate that improvements were broadly distributed rather than driven by a small subset of cases.

\section{Discussion}

In this study, we introduced a physics-informed hybrid CNN--diffusion framework that learns to invert susceptibility-induced ssEPI distortions by coupling realistic forward physical simulation with anatomically guided neural reconstruction. Rather than relying on explicit distortion field estimation or additional acquisitions, the proposed approach leverages large-scale simulated distorted/undistorted pairs and T2-weighted anatomical conditioning to enable robust correction under severe off-resonance conditions. Across both synthetic benchmarks and challenging clinical cases, this strategy demonstrated consistent improvements in geometric fidelity and interpretability, including scenarios where conventional field-map--based or reverse phase-encoding methods are known to break down. These findings suggest that learning an approximate inverse of a realistic forward distortion model provides a viable alternative to acquisition-dependent correction pipelines in prostate DWI.

At a mechanistic level, DGR differs from conventional EPI correction in that it does not depend on estimating a single distortion field. Instead, it learns a data-driven restoration that jointly leverages diffusion contrast (low-b DWI and ADC) and an aligned T2-weighted image as an anatomical constraint, which is particularly valuable in regions where pile-up or signal voids make analytical inversion ill-posed. Architecturally, the CNN front-end provides a physics-consistent coarse correction of global geometry, while the diffusion stage performs conservative refinement to recover local texture and more realistic noise behavior (most noticeably on ADC). Finally, training on a broad set of simulated distortions derived from measured B$_0$ maps exposes the model to diverse nonlinear deformations, helping it handle stretching, compression, and localized signal loss that are difficult to capture with purely analytical models.

Several limitations should be acknowledged. First, the forward simulator was driven primarily by B$_0$ templates measured in hip-implant patients. Although we applied extensive perturbations to increase diversity, susceptibility patterns caused by rectal distension---characterized by milder, more gradual geometric changes---may be only partially represented in our simulated distortions. Expanding the template library to include rectal-distension cases will be important directions for future work.

Second, while the diffusion module was used conservatively (img2img strength = 0.1) to minimize the risk of hallucination, generative models inherently carry the possibility of introducing subtle structures that may not reflect true signal. Although no false lesions were observed in our qualitative evaluation, larger-scale studies will be necessary to ensure that diffusion-based refinement does not inadvertently alter clinically relevant features.

\section{Conclusion}

We introduced a physics-informed learning framework that restores prostate DWI distorted by ssEPI through a hybrid CNN--diffusion architecture trained on large-scale, simulated forward models. Across synthetic benchmarks and real clinical cases, the method substantially improves distortion recovery and diagnostic interpretability, outperforming FSL FUGUE and TOPUP without requiring additional acquisitions such as B$_0$ field maps or reverse phase-encoding scans. By learning a robust inverse mapping grounded in physical distortion mechanisms, the proposed DGR framework offers a practical and scalable solution for improving prostate DWI quality, particularly in metal-affected imaging where conventional methods often fail.

\section*{Code Availability}

To support reproducibility and accelerate research use, the DGR codebase will be released at \url{https://github.com/Albertlongzi/DGR}.

\section*{Acknowledgments}

We extend our gratitude to the Research Imaging Core (RIC) at Cedars-Sinai for their valuable support. Special thanks to MRI Technologist Mike Ngo and Irene Lee for scan sequence instruction, and Nurses Catherine Ubaldo-Prado and Lee Hyae for their assistance in patient setup.

\bibliographystyle{unsrt}  
\bibliography{references}

\end{document}